\documentclass{epl}

\usepackage{psfig}

\title{Classification of three-body quantum halos}

\author{A.S.~Jensen\inst{1} \and K.~Riisager\inst{1} \and 
          D.V.~Fedorov\inst{1} \and E.~Garrido\inst{2}}
\institute{
  \inst{1} Institute of Physics and Astronomy, Aarhus University -
                                            DK-8000 Aarhus C, Denmark \\
  \inst{2} Instituto de Estructura de la Materia, CSIC - Serrano 123,
                                            E-28006 Madrid, Spain
}

\begin{document}

\maketitle

\begin{abstract} 
The different kinds of behaviour of three-body systems in the weak
binding limit are classified with specific attention to the transition
from a true three-body system to an effective two-body system.  For
weakly bound Borromean systems approaching the limit of binding we
show that the size-binding energy relation is an almost universal
function of the three $s$-wave scattering lengths measured in units of
a hyperradial scaling parameter defined as a mass weighted average of
two-body equivalent square well radii.  We explain why three-body
halos follow this curve and why systems appearing above reveal
two-body substructures.  Three-body quantum halos 2-3 times larger
than the limit set by zero hypermoment are possible.
\end{abstract}

\section{Introduction}

Attempts for a general classification of halo states were started
early in the development of the field, see e.g.\ \cite{rii94}, and
have recently led to the suggested definition of halo states
\cite{rii00,jen00} as having more than 50\% probability of being in a
cluster configuration where more than 50\% of this probability should
be in a classically forbidden region.  This definition is
straightforward to apply for a two-body system, where one basically
has to find the outer classical turning point in the radial motion.
The three-body systems are more challenging \cite{nie01} and it is the
purpose of this paper to discuss their possible modes of behaviour.

One obvious point that needs clarification is how to generalize the
outer classical turning point, that also can be used \cite{rii00} to
scale different halo systems so that e.g. nuclear and molecular halos
can be compared in dimensionless units.  For the interesting special
case of Efimov states a universal scaling property predicting one
Efimov state from the previous one has been developed, see
\cite{efi70,amo97,fre99}.  Another point that has only been discussed
briefly in the literature so far is how the transition from a
three-body to a two-body state takes place as the binding potentials
are changed \cite{fed94}.  Connected to this is the classification of
possible three-body configurations into Borromean
\cite{zhu93,goy95,mos00}, tango \cite{rob99} or other bound states.
To clarify the principles we shall mainly consider systems where all
particles are in relative $s$-waves, which dominate at large
distances.  Results of more realistic calculations will also be given
for $^{11}$Li and the hypertriton.

\section{Three-body systems}

There is naturally more variability in three-body systems than in
two-body systems.  The three two-body subsystems might all play a role
in the asymptotic region, so even in the weak-binding limit we should
expect several types of behaviour to be possible, even for the
simplest case of only relative $s$-waves.  Systems with zero and one
bound subsystem are called Borromean and tango systems, respectively.

In a two-body system the classical turning points are found by
equating the total energy and the potential energy.  In principle we
can generalize this to three particles in a specified quantum state
described by the wave function $\Psi$ and find the probability for
being in the non-classical region as $\int|\Psi|^2 {\rm d} \tau$,
where the integration is confined to regions with potential energy
larger than the total energy.  This will be much harder to calculate
than for a two-body system and could therefore be a rather impractical
condition.  Furthermore, we are not assured that the wavefunction will
behave in a simple way in the non-classical region; there might be
configurations where one pair is in a forbidden region whereas the
third particle is in an allowed region.  We shall therefore here
rather explore the possibility of generalizing the classical turning
point into a three-body scaling radius and discuss two different types
of ``derivation'' of it.

\section{First derivation}

We use hyperspherical coordinates to describe the relative motion of
three particles with masses $m_i$, where $i=1,2,3$. The total mass is
$M = m_1 + m_2 + m_3$, the individual momenta and coordinates are
${\bf p_i}$ and ${\bf r_i}$ and the hyperradius $\rho$ is defined by
\begin{equation} \label{e20}
 m \rho^2 \equiv  \frac{1}{ M}  \sum_{i<k} m_i m_k 
({\bf r_i} - {\bf r_k})^2  = \sum_i m_i ({\bf r_i} - {\bf R})^2  \;  ,
\end{equation}
where ${\bf R}$ is the center of mass coordinate and $m$ is a mass
unit chosen for convenience.  The hyperradius is an average radius
coordinate, applicable to all three-body systems and useful for all
angular momenta and for non-spherical systems.  The total mean square
radius $\langle r^2 \rangle$ is then via the particle sizes $\langle
r^2 \rangle_i$ given by
\begin{equation} \label{e21}
 M \langle r^2 \rangle = m \langle \rho^2 \rangle +
      \sum_i m_i \langle r^2 \rangle_i  \; .
\end{equation}
It is natural to choose a three-body scaling radius $\rho_0$ so that
the arbitrary mass $m$ enters in the same way in $\rho$ and $\rho_0$
and all measures of size, that typically rely on their ratio, become
independent of $m$.  The two-body scaling property relating size and
binding energy can then be generalized if $\langle \rho^2 / \rho_0^2
\rangle$ is an almost single-valued function of another dimensionless
quantities $B m \rho^2_0/ \hbar^2$, where $B$ is the three-body
binding energy \cite{fed94}.  Such a scaling property is clearly an
advantage when searching for a general definition of the scaling
radius $\rho_0$.  The relation should apply for systems consisting of
particles with widely different masses and ranges of interactions.  A
tempting definition of $\rho_0$ is to maintain the complete analogy to
$\rho$, i.e.
\begin{equation} \label{e22}
 m \rho^2_0 \equiv  \frac{1}{ M}  \sum_{i<k} m_i m_k R_{ik}^2  \;  ,
\end{equation}
where $R_{ik}$ is interpreted as the equivalent square well radius of
the system consisting of particle $i$ and $k$.  As argued in
\cite{rii00} this definition is convenient in descriptions of
three-body systems intermediate between two and three-body scaling.

\section{Hyperradial potential}

The choice of hyperspherical coordinates leads to effective radial
potentials obtained by adiabatic expansion or by averaging in other
ways over the remaining set of angular coordinates.  The classically
allowed regions for such one-dimensional potentials are easily
defined. However, they could be completely different from the regions
where the three pairs of particles are located in their classically
allowed regions defined by the corresponding two-body potentials.  In
fact, it is entirely possible to have classical motion in the
hyperradial coordinate while the system is in non-classical regions in
real space. Hyperradial turning points are therefore useless as
definitions of quantum halos and often without any resemblance to the
length unit $\rho_0$ defined above in terms of two-body properties.

It is instructive to consider the general behaviour of the hyperradial
potential. For zero-range two-body potentials the only energy
available through combination of parameters is $\hbar^2/(2 \mu
\rho^2)$, where $\mu$ is a combination of reduced masses. With this
large distance behaviour any number of solutions is possible, ranging
from zero to the infinitely many Efimov states.  For finite range
interactions the ranges or alternatively the scattering lengths
provide additional length parameters and the hyperradial potentials
could approach zero faster than $\rho^{-2}$.

From \cite{jen97} we obtain the large distance behaviour of the
dominating lowest $s$-wave adiabatic potential as
\begin{equation} \label{e66}
V_{ad} = - \frac{\hbar^2}{2 m \rho^2}  \frac{16}{\pi}
\sum_{i<k} \frac{a_{ik}}{\rho} \sqrt{\frac{\mu_{ik}} {m}}  \equiv 
\frac{\hbar^2}{2 m \rho^2}  \frac{48}{\pi \sqrt{2}} \frac{a_{av}}{\rho }  \;  ,
\end{equation}
where the reduced mass is $\mu_{ik}= m_i m_k/(m_i+m_k)$, the $s$-wave
scattering length is $a_{ik}$ and the average scattering length is
defined as
\begin{equation} \label{e76}
a_{av} \sqrt{m} \equiv 
\frac{\sqrt{2}}{3} \sum_{i<k} \sqrt{\mu_{ik}} a_{ik} \; . 
\end{equation}
The chosen normalization reduces $a_{av}$ to the common $a_{ik}$ when
all three particles are identical.  Thus the large distance behavior
of $V_{ad}$ in Eq.(\ref{e66}) is $\rho^{-3}$ which is reached when
$\rho$ is comparable to $ 48 a_{av} / (\pi \sqrt{2})$, see
\cite{jen97}.

\section{Second derivation}

Scaling can be shown analytically to occur for the special case of a
$K=0$ wavefunction ($K$ being the hypermoment) for square well two-body
potentials of depth $S_{ik}$ and radius $R_{ik}$.  Here the effective
hyperradial potential $V$ can be obtained \cite{fed94} as
\begin{equation} \label{e36}
V(\rho) = \frac{16}{3 \pi} \sum_{i<k} \frac{S_{ik}}{\rho^3}
 \Big(\sqrt{\frac{\mu_{ik}}{m}} R_{ik}\Big)^3 \; ,
\end{equation}
which is valid when $\rho$ is several times larger than any of the
square well radii.  The square well two-body $s$-wave scattering
length is given by $a_{ik}/R_{ik} = -1 +
\tan{(K_{ik}R_{ik})}/(K_{ik}R_{ik})$, where $K_{ik}$ is the zero
energy wave number inside the square well, i.e $\hbar^2 K_{ik}^2 /(2
\mu_{ik}) = S_{ik}$.  Then $K_{ik}^2 R_{ik}^2 = 2 S_{ik} \mu_{ik}
R_{ik}^2 / \hbar^2$ is a specific function of $a_{ik}/R_{ik}$, which
approaches the constant $\pi^2 /4$ when $a_{ik}$ becomes much larger
than $R_{ik}$.  Thus the effective radial potential in Eq.(\ref{e36})
approaches the form
\begin{equation} \label{e46}
V(\rho) =  \frac{\hbar^2}{2 m \rho^2} \frac{4 \pi}{3}
\sum_{i<k} \frac{R_{ik}}{\rho} \sqrt{\frac{\mu_{ik}}{m}}  \equiv 
\frac{\hbar^2}{2 m \rho^2} \sqrt{8} \pi  \frac{\rho_0}{\rho}  \;  ,
\end{equation}
where the definition of $\rho_0$,
\begin{equation} \label{e56}
\rho_0 \sqrt{m} \equiv \frac{\sqrt{2}}{3} \sum_{i<k} \sqrt{\mu_{ik}}R_{ik} \;,
\end{equation}
reduces to that of Eq.(\ref{e22}) for identical masses and radii.
Note that Eq.(\ref{e56}), in contrast to Eq.(\ref{e22}), employs a
linear (not squared) summation.  For systems where the two-body
scattering lengths are large, most of the wavefunction will reside in
the region where Eq.(\ref{e46}) holds for $K=0$ and one clearly has
scaling.

A desirable property of the scaling radius $\rho_0$ would be that
classically forbidden regions on average are given roughly by $\rho >
\rho_0$.  We can consider this question briefly for the specific $K=0$
solutions, where the probability $P_{ik}$ for particles $i$ and $k$
being inside their square well radius $R_{ik}$ for large $\rho$ is
\begin{equation} \label{e86}
P_{ik}(\rho) = \frac{16}{3 \pi} \Big(
\frac{R_{ik}}{\rho} \sqrt{\frac{\mu_{ik}}{m}} \Big) ^3 \; .
\end{equation}
This probability is 1/2 for $\rho = 2(4/(3\pi))^{1/3}
\sqrt{\mu_{ik}/m}R_{ik}$ which indicates that systems reside mainly in
classically forbidden regions if their mean square radii are somewhat
larger than $\rho_0^2$.

We do not believe that the $K=0$ wave functions are realistic
solutions for the weakly bound systems since they do not allow for any
form of correlations between the particles.  For very loosely bound
systems where $a_{av}$ is much larger than $\rho_0$ the potential will
fall off as $\rho^{-2}$ between $\rho_0$ and $a_{av}$, i.e.\ slower
than $\rho^{-3}$ as for a pure $K=0$ solution.

\begin{figure}[!t]
\centerline{\psfig{figure=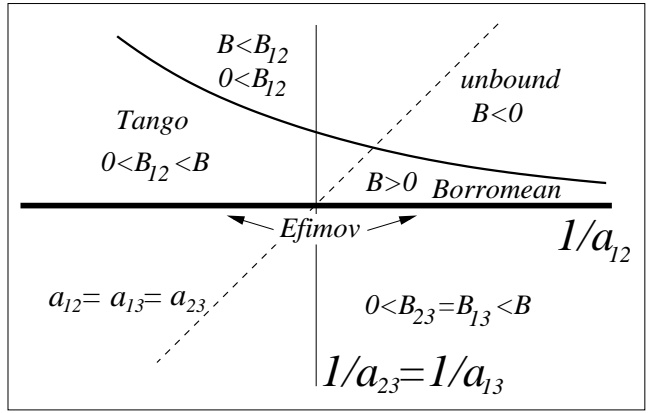,width=8.5cm}}
\vspace*{0.2cm}
\caption{Sketch of the different regions of stability for a three-body
system as functions of the inverse $s$-wave two-body scattering
lengths $a_{ik}$.  The central point corresponds to all $a_{ik} =
\infty$ which is assumed to be the threshold for binding of the first
state in the $i-k$ subsystem.  All potentials are attractive or
vanishing. Particles $1$ and $2$ are identical. The dashed line
represents three identical particles. The two and three-body binding
energies are $B_{ik}$ and $B$, respectively.}
\label{fig1}
\end{figure}

\section{Scaling properties}

The scaling radius will be used to look for scaling properties of
three-body systems in the weak binding limit.  We are aiming for as
universal properties as possible, but the rather different types of
structure that occur in three-body systems means we first have to look
at how they can be classified.  In fig.\ref{fig1} we illustrate the
various stability regions as function of scattering lengths.  At the
border between tango and Borromean regions one subsystem has a bound
state with zero energy. On the thick horizontal line two identical
subsystems have bound states of zero energy and the infinitely many
Efimov states arise.

\begin{figure}[!t]
\centerline{\psfig{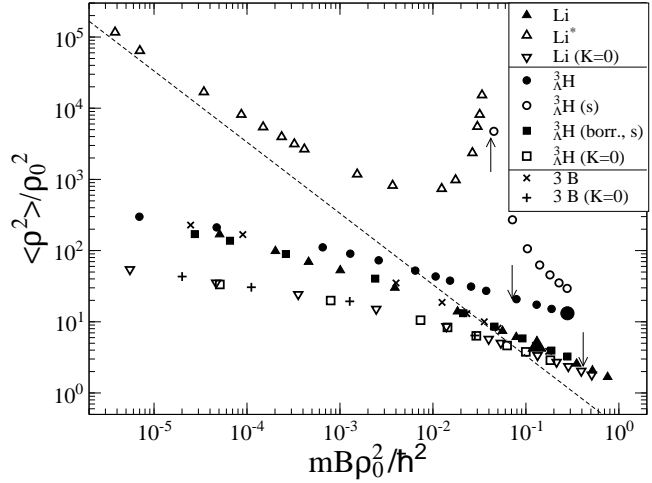}}
\vspace*{0.0cm}
\caption{Scaling plot for three-body halos.  The ratio of the halo and
effective potential mean square radii is plotted versus the scaled
separation energy.  The definition of scaling radius in
Eq.(\protect\ref{e56}) is used.  The dashed line represents the Efimov
states for minimum attraction. Triangles are for masses corresponding
to $^{11}$Li. Squares and circles are for $^{3}_{\Lambda}$H. The
realistic points are indicated by a large closed triangle and square,
respectively.  The stars and crosses refer to a system of three
different particles with two fixed scattering lengths while the third
is varied.  Almost indistinguishable curves arise for three identical
Borromean bosons by varying the common scattering length.  The arrows
indicate transitions from (i) Borromean to tango region (closed
circles), (ii) tango directly to unbound three-body system (open
circles), (iii) Borromean to either two bound subsystems or the tango
region (stars, crosses, open and closed triangles and squares) all
occurring approximately at the same point. The excited states of
$^{11}$Li all correspond to a bound n-$^{9}$Li system.}
\label{fig2}
\end{figure}

The figure only shows the region where $a_{23}=a_{13}$, appropriate
e.g.\ when particles 1 and 2 are identical, but already indicates that
two distinct types of transitions can occur as a function of $a_{12}$,
namely moving from the tango region (the 23 and 13 subsystems are
unbound) from left to right either directly into unbound systems or
through the Borromean region.  To see this in detail we now turn to
the numerical results and show in fig.~\ref{fig2} the region of weak
binding for a number of both schematic and realistic examples.
Through Eq.(\ref{e21}) this scaling plot displays the mean square
radius of the system in units of $\rho_0^2$ versus the dimensionless
binding energy $mB\rho_0^2/\hbar^2$ \cite{rii00}, where $B$ is the
three-body binding energy. Zero binding energy corresponds to three
non-interacting particles at rest.

We shall in this letter use the scaling radius defined in
Eq.(\ref{e56}), where $R_{ik}$ is the radius of the square well with
the same scattering length and effective range as the actual two-body
potential.  Changing to the other definition in Eq.(\ref{e22}) has
essentially no effect on the hypertriton examples, whereas the
different $^{11}$Li points move by about 30\%.  The resulting figure
is practically indistinguishable from the present one, except that the
perfect scaling for $K=0$ is violated slightly.  There is no strong
preference for any of these two definitions of $\rho_0$.

The first analysis in \cite{fed94} assumed that simple systems
at least aymptotically would correspond to a single value of the
hypermoment $K$.  We have artificially restricted the wavefunctions
for different systems to contain only $K$=0 terms and find that these
systems, as argued above, indeed do scale and lie on a single curve.
However, the points corresponding to more realistic calculations lie
above this $K$=0 curve and can be grouped and understood as follows.

In the realistic $^{3}_{\Lambda}$H-structure the neutron and proton
are almost in a deuteron configuration (close circles) whereas the
${\Lambda}$-particle is far outside but still bound by about
0.14~MeV. Thus this is a tango system and the large size is almost
entirely due to the ${\Lambda}$-deuteron extension, i.e. of two-body
character.  By using a simplified form for the neutron-proton
interaction with only $s$-waves the position is sligthly higher than
the realistic point.  By decreasing this neutron-proton attraction the
binding decreases moderately while the radius drastically increases
(open circles). This off hand surprising property can be understood
from fig.~\ref{fig1} by moving horizontally from the tango region to
the right, increasing $1/a_{12}$. For sufficiently weak initial
binding, i.e.\ large values of $1/a_{13}$, the threshold for
${\Lambda}$-deuteron binding is reached instead of the Borromean
region as we decrease the attraction of the only bound subsystem. Thus
the diverging radius is due to ${\Lambda}$-deuteron two-body threshold
and not related to the neutron-proton threshold.

Decreasing the $s$-wave attraction between neutron and proton while
maintaining all other parts of the realistic interaction (filled
circles), now leads from tango into the Borromean region of
fig.\ref{fig1}, i.e. the three-body system remains bound even after
the neutron-proton potential is too weak to form a bound two-body
state. The result is a completely different curve in fig.~\ref{fig2}
much more in agreement with the logarithmic divergence expected from
\cite{fed94}, although still significantly above the $K=0$ curve.  We
stress that nothing drastically happens at the arrow where the
deuteron becomes unbound (scaled binding energy about 0.08).

The realistic point for the Borromean nucleus $^{11}$Li is at a scaled
binding energy of 0.13 in fig.~\ref{fig2}.  By decreasing the $s$-wave
attraction in the neutron-$^{9}$Li systems, i.e.\ going vertically
upwards in fig.~\ref{fig1} approaching the threshold, the binding
decreases and the radius increases corresponding to a logarithmically
diverging curve in fig.~\ref{fig2} (filled triangles) eventually
approaching that of the Borromean hypertriton example.  A similar
behaviour is seen for a hypothetical hypertriton with only $s$-waves
included (filled squares) but with a slightly increased
nucleon-$\Lambda$ attraction. Then we move horizontally from the tango
to the Borromean region.  For a Borromean system of three particles
(stars) where the three masses and scattering lengths are equal or
differ substantially, we again follow the same trajectory by changing
one scattering length.

By increasing the $s$-wave attraction in the neutron-$^9$Li systems,
i.e.\ vertically approaching the Efimov limit in fig.~\ref{fig1},
larger binding and smaller radius result as expected. The resulting
non-Borromean $^{11}$Li ground state resembles more and more an
ordinary nuclear state.  However, at some point an excited state
appears, i.e.\ the first Efimov state (open triangles). At first it is
very weakly bound and close to the Efimov line.  As the attraction and
the three-body binding energy increases the size at first decrease and
then ``turn around'' and increase again as the binding energy of the
two bound two-body subsystems approach and finally overtake the
three-body binding energy \cite{efi70}.  As the two-body threshold is
approached the third excited state (second Efimov state) should appear
on the dashed line. Unfortunately the mass ratio (neutron to $^{9}$Li)
is relatively small and the next state is many orders of magnitude
outside the scale of the plot, i.e. outside the reach of experimental
as well as most numerical techniques.

One remaining question is the approach of the two sets of points,
related to hypertriton and $^{11}$Li, at very small binding
energy. Both these sets arise by approach of (different) thresholds
for Borromean binding in fig.~\ref{fig1}.  When the majority of the
radial wave function is located in the tail of the adiabatic potential
in Eq.(\ref{e66}) then $a_{av}$ is a decisive length parameter. One
could then erroneously be led to conclude that $a_{av}$ determines the
size-binding relations and the threshold for Borromean binding in
fig.~\ref{fig1}.  However, the absolute scale of the energy can not be
determined from one parameter alone, the short distance behaviour is
also indispensably necessary \cite{amo97,fre99,fed01}.

To understand this we imagine that we reduce the ranges of the
potentials while increasing the strengths to maintain the scattering
lengths. This zero-range limit results in the Thomas collapse and
infinitely many bound states at small distances \cite{fed01}.  To
avoid this unphysical behaviour some kind of renormalization is
needed. The simplest is to maintain the large distance behavior while
only using the two-body potentials down to a distance below which the
structure is uninteresting. The two-body results can then be expressed
in units of such a rather arbitrary length parameter. The effect on a
three-body system can then be anticipated mimicked by a similar
renormalization by use of a hyperspherical length unit.  However, this
is precisely the content of fig.~\ref{fig2} where we used $\rho_0$ as
the scale parameter.

The curves in fig.~\ref{fig2} coinciding at small binding are
therefore an almost universal curve as indicated by the convergence in
the figure.  It would perhaps be rather fortuitous if this simple
renormalization procedure in hyperradius results in a universal
rescaled curve. The average scattering lengths for the four cases in
the weak binding limit vary from 4~fm to 20~fm whereas the ratio
$a_{av}/\rho_0$ varies between 1.7 and 4.2. However, $a_{av}$ only
determines the adiabatic potential for distances smaller than the
scattering lengths and larger than the potential ranges. These
conditions are not well fulfilled for the examples in fig.~\ref{fig2}.
Thus constant $a_{av}/\rho_0$ should not necessarily arise, since
$a_{av}$ should be replaced by a complicated function of all three
$a_{ik}$. This is not contradicting the numerical results in
fig.~\ref{fig2} which is obtained without use of $a_{av}$.  Thus the
emerging numerical curve is almost universal providing rather well
defined scaling properties.

\section{Summary and conclusion}

Three-body quantum halos were previously believed to appear along the
$K=0$ curve \cite{fed94}. This conclusion was reached by omitting
constant terms compared to the leading order logarithmically diverging
term.  However, the present more refined analysis reveals the correct
higher lying universal curve approached in the weak binding limit. In
fact asymptotically these curves differ by a constant factor. In any
case three-body quantum halos can appear above the $K=0$ curve,
i.e. allowed in a larger window and being a factor of 2--3 larger than
expected.  One implication is that $K=0$ wavefunctions cannot be used
in the weak binding limit due to the coherent large-distance
contributions from more than one subsystem. For stronger binding a
$K=0$ basis is simply incomplete.

In fig.~\ref{fig1} ground state three-body halos appear just below the
line separating the Borromean from the unbound region.  On top of this
the Efimov effect can also give ``super-halos'' in excited states.
Ground state two-body halos can appear just below the line separating
the tango from the unbound region.  In fig.~\ref{fig2}, three-body
halos where the three-body dynamics dominate will appear on or below
the (almost) universal curve.  All systems falling above are
influenced by an effective two-body threshold and correspond either to
the Efimov line (caused by the threshold in two subsystems) or the
tango-unbound line in fig.~\ref{fig1}.  We stress that all quantities
used to place a system in fig.~\ref{fig2} in principle can be measured
experimentally and that our results therefore can be used to classify
the three-body nature of realistic systems.

\end{document}